\documentclass[a4paper]{article}
\usepackage{amsmath}
\usepackage{amssymb}
\newcommand{\haak}[1]{\left(#1\right)}

\newcommand{\lhaak}[1]{\left | #1\right |}

\newcommand{\gem}[1]{\left\langle #1\right\rangle}

\newcommand{\lhaakr}[1]{\left.#1\right |}

\begin{document}

\begin{titlepage}
\rightline{\large May 2003}
\vskip 2cm
\centerline{\Large \bf Detecting mirror matter on Earth}
\centerline{\Large \bf via its thermal imprint
on ordinary matter}
\vskip 1.2cm
\centerline{\large R. Foot$^{a}$ and S. Mitra$^{b}$\footnote{
E-mail address: foot@physics.unimelb.edu.au, saibalm@science.uva.nl}
}

\vskip 0.7cm
\centerline{\large \it $^{a}$ School of Physics,}
\centerline{\large \it University of Melbourne,}
\centerline{\large \it Victoria 3010 Australia}
\vskip 0.5cm
\centerline{\large \it $^{b}$ Instituut voor Theoretische Fysica,}
\centerline{\large \it Unversiteit van Amsterdam, 1018 XE Amsterdam,}
\centerline{\large \it The Netherlands}

\vskip 2cm

\noindent
Mirror matter type dark matter can exist on the Earth's surface,
potentially in enhanced concentrations at various anomalous impact
sites. Mirror matter fragments can draw in heat from the 
ordinary matter environment, radiate 
mirror photons and thereby cool the surrounding
ordinary matter. We quantify this effect and suggest that it could
be used to help locate mirror matter deposits.
This method, together with the centrifuge technique, seems to
provide the most promising means to experimentally detect mirror matter
type dark matter in the Earth.

\end{titlepage}

One of the most fascinating candidates for the non-baryonic
dark matter in the Universe is mirror matter\cite{ly,flv,blin}. From a 
particle physics point of view, mirror matter is a natural
candidate for dark matter because:
\begin{itemize}
\item
It is theoretically well motivated since it is predicted 
to exist if parity and
time reversal symmetries are exact, 
unbroken symmetries of nature.

\item
It is necessarily dark and stable: Mirror baryons have
the same lifetime as ordinary baryons and couple to
mirror photons instead of ordinary photons.

\item
Recent observations\cite{wmap} suggest that the cosmic abundance of
non-baryonic dark matter is of the same order of magnitude as
ordinary matter
$\Omega_{\text{b}} \sim \Omega_{\text{dark}}$. A result which can
naturally occur if dark matter is identified with mirror matter \cite{dark}.

\end{itemize}

It is important to realize that mirror matter fragments can
interact with ordinary matter non-gravitationally via the
photon-mirror photon kinetic mixing interaction \cite{flv,km}:
\begin{eqnarray}
\mathcal{L} = \frac{\epsilon}{2}
F^{\mu \nu} F'_{\mu \nu},
\end{eqnarray}
where $F^{\mu \nu}$ ($F'_{\mu \nu}$) is the field strength 
tensor for electromagnetism (mirror electromagnetism).
One effect of the above interaction is that it gives the mirror
charged particles a small electric charge, $\epsilon e$ \cite{km,flv}.

If there is some amount of mirror matter in the solar system
then the photon-mirror photon kinetic mixing interaction will cause
unique effects when mirror matter space-bodies collide
with the planets/asteroids \cite{footyoon1,footyoon2,eros} and also when 
ordinary spacecraft propagate
out through the solar system \cite{fvpioneer}. 
Motivation to study such effects comes
from observations of anomalous (Earth) impact events on both
small and large scales \cite{small,tunguska,haines}, as well as the
observed anomalous slow-down of {\it both} Pioneer spacecraft \cite{study}.
Also, the remarkable observations of the NEAR Shoemaker 
space-probe \cite{near1} also seem to support the mirror matter
hypothesis.

It turns out that\cite{eros} these observations can be explained
if the photon-mirror photon kinetic mixing is in the range: 
\begin{eqnarray}
10^{-9} \lesssim \lhaak{\epsilon} \lesssim 10^{-6}.
\label{zzz}
\end{eqnarray}
Interestingly, if 
$\epsilon$ is in the above range, mirror matter should exist
on the Earth's surface since the photon-mirror photon kinetic mixing
interaction is then strong enough to oppose the force 
of gravity\cite{centrifuge}.
This means that  mirror matter can
potentially be experimentally discovered in the ground, at various
anomalous impact sites.

One way to detect the existence of mirror
matter in a sample of earth is to use a centrifuge\cite{centrifuge}. 
Centrifuging a sample containing mirror matter will expel mirror
matter fragments thereby leading to a potentially observable weight loss
of the sample.  
Another complementary way - the subject of the present paper --
is to search for its thermal imprint.
The idea is that a piece of mirror matter in the ground would
draw in heat from the surrounding ordinary matter and radiate
it away as mirror photons, thereby cooling the surrounding ordinary
matter. 

This idea was mentioned earlier in Ref.\cite{footyoon2} where the temperature
imprint on the surrounding ordinary matter was evaluated
in terms of the surface temperature of the mirror body.
Thus, an important issue is to calculate the temperature of 
a mirror matter body embedded in ordinary matter.
As we will show in this paper, 
for $\epsilon$ in the entire range of interest, Eq.(\ref{zzz}),
and assuming Earth-like temperatures,
the interaction between ordinary and mirror matter is
strong enough for thermal equilibration to occur. 
This result is quite important, since it means that
any significant mirror matter deposit could be detected
by its thermal imprint on the surrounding ordinary matter.

Consider a mirror matter fragment at temperature $T'$ embedded within 
(solid) ordinary matter at temperature $T>T'$. 
[Mirror matter would necessarily be embedded {\it within} ordinary matter
if $\epsilon < 0$
since the electrostatic force between ordinary and mirror
matter is attractive \cite{footyoon2}].
The mirror atoms in the mirror matter fragment undergo 
frequent collisions with ordinary atoms and,
on average, gain energy. 
If the temperature is not too 
low\footnote{For temperatures much lower than
the Debye temperature, less than room temperature for most solids, the
high frequency vibrational modes of the solid become frozen into the 
ground state, and classical statistical
mechanics breaks down.}, classical statistical mechanics
can be applied to calculate the heat transfer.

The heat transfer per unit volume, $dQ/dV$, can be expressed as the product
(appropriately averaged over the Maxwellian
velocity distributions)
of the collision rate $\Gamma$, number density $n'$ and energy 
transfer per collision $\Delta E_{\text{kin}}$:
\begin{eqnarray}
\frac{dQ}{dV} &=& \langle \Gamma n' \Delta E_{\text{kin}} \rangle \nonumber \\
&=& \langle \sigma v_{\text{rel}} n n' \Delta E_{\text{kin}} \rangle .
\label{mon1}
\end{eqnarray}
In Ref.\cite{centrifuge} it was shown
that for typical thermal velocities at room temperature the 
cross section is approximately isotropic and given by:
\begin{eqnarray}
\sigma &\sim & 10^{-2} \epsilon^2 \text{ cm}^2 \text{ for } |\epsilon |
\lesssim 10^{-8} \nonumber \\
\sigma &\sim & 10^{-18}\text{ cm}^2 \text{ for } |\epsilon | \gtrsim
10^{-8}. \label{sigma}
\end{eqnarray}
Since the collisions are approximately isotropic,
the energy transfer per collision is roughly:
\begin{eqnarray}
\langle \Delta E_{\text{kin}} \rangle \sim \frac{m}{2} (v^2 - v'^2)
\sim \frac{3}{2} k(T-T'),
\end{eqnarray} 
where we assume for simplicity that the ordinary and mirror material is
composed of atoms of mass $m$.
Using the above, Eq.(\ref{mon1}) can be
solved (a more precise computation is given in the appendix)
leading to:
\begin{equation}\label{result}
\frac{dQ}{dV}=
4\pi^{-1/2}m^{-1/2}k^{3/2}\sigma
n'n\haak{\frac{T'+T}{2}}^{1/2}\haak{T-T'} .
\end{equation}
This is the rate
which heat is transferred from ordinary matter to mirror matter.
We now consider the effect of this on a fragment of mirror matter
embedded in ordinary matter.

Consider a spherical mirror matter `rock', of radius $R'$
completely embedded within ordinary matter.
If its surface temperature is $T'_{R'}$, then it
will emit mirror photons, with energy
loss $\mu \sigma {T'}_{R'}^4 4\pi R'^2$ (where
$\mu \le 1$ is the emissivity of the surface). Meanwhile, energy is absorbed
from ordinary matter. The energy absorbed 
depends on the rate of energy transfer, $\frac{dQ}{dV}$, so that
\begin{eqnarray}
\int^{R'}_{0}\frac{dQ}{dV} 4\pi r^2 dr = \mu \sigma {T'}_{R'}^4 4\pi
R'^2 .
\label{1}
\end{eqnarray}
Putting the numbers in (taking $m \sim 20 m_{\text{proton}}$,
$n \sim n' \sim 10^{23}/\text{cm}^3$)
and defining $u \equiv (T-T')/T$,
we find:
\begin{eqnarray}
\frac{dQ}{dV} \approx  \Lambda 
\haak{\frac{T}{300 K}}^{3/2} \haak{1-\frac{u}{2}}^{1/2}u \ 3\times
10^{12}\ \text{Joules/}(\text{cm}^{3}\text{.s})
\end{eqnarray}
where $\Lambda = \haak{ \frac{\epsilon}{10^{-8}} }^2
$ for $|\epsilon | \lesssim 10^{-8}$ and
$\Lambda = 1$ for $|\epsilon | \gtrsim 10^{-8}$.
Thus, Eq.(\ref{1}) becomes (after dividing both sides of the equation by
$R'^2$):
\begin{eqnarray}
\Lambda
\left( \frac{R'}{\text{cm}}\right)
\gem{u\haak{1-\frac{u}{2}}^{1/2}}
\left( \frac{T}{300 K}\right)^{3/2}
10^{13}
= 0.6 \mu \left( \frac{{T'}_{R'}}{300 K}\right)^4 , 
\label{3}
\end{eqnarray}
where we have defined the average of function $f(r)$ inside 
the region $r<R'$ as 
\begin{eqnarray}
\gem{f} \equiv 
\frac{3}{R'^3}\int^{R'}_0 f(r) r^2 dr .
\end{eqnarray}
Clearly, Eq.(\ref{3}) shows that for an Earth-like environment ($T \sim
300 K$),
$|\epsilon | \gtrsim 10^{-12}$
[which includes the entire range of most interest, Eq.(\ref{zzz})]
and for all $R'$ of interest (including microparticle sizes, $R' \sim
10^{-4}\text{ cm}$),
$\gem{u}\ll 1$.
This implies that
\begin{eqnarray}
\gem{T'} \simeq T
\end{eqnarray}
which means that the mirror body thermally equilibrates with its 
ordinary matter surroundings.

If the mirror thermal conductivity ($\kappa'$)
were very poor,
one could still have ${T'}_{R'} \ll T$ (which is technically the
case for $\kappa' \to 0$).
However, for an Earth-like environment, this effect is negligible.
To see this,
consider a spherical shell $r_1 < r < R'$. In a steady-state situation,
the energy going into this shell must equal the energy going out,
which translates into the following equation:
\begin{eqnarray}
\int^{R'}_{r_1} \frac{dQ}{dV} 4\pi r^2 dr - 
\kappa' \frac{\partial T'}{\partial r} 
4\pi r^2_1 = \mu \sigma {T'}_{R'}^4 4\pi R'^2 .
\end{eqnarray}
[Of course this equation reduces to the earlier one, Eq.(\ref{1}),
for $r_1 = 0$].
For $r_1$ sufficiently close to $R'$, the thermal
conductivity term (i.e. second term on the left-hand side
of the above equation) dominates over the first term.
In this case, we can evaluate the mirror temperature
gradient near the surface of the mirror body:
\begin{eqnarray}
\lhaakr{\frac{\partial T'}{\partial r}}_{r=R'} \approx  
-3 \haak{ \frac{{T'}_{R'}}{300 K}}^4 
\haak{ \frac{0.016 J/(K.s.cm)}{\kappa'}}
K/\text{cm}
\end{eqnarray}
where, for convenience, we have compared $\kappa'$ to the
average value of $\kappa$ for the Earth's crust
($\langle \kappa \rangle \approx  0.016 \text{ J/K.s.cm}$).
This is a very small gradient (for ${T'}_{R'} \lesssim 300 K$)
which means that the mirror thermal conduction term is
negligible in comparison to the ordinary-mirror matter energy
transfer term.

The conclusion is, that for Earth-like temperatures ($T
\lesssim 300 K$), a mirror matter body would
equilibrate in temperature with the ordinary matter 
surroundings and emit mirror radiation from its
surface, thereby cooling the ordinary matter.

The amount of radiation emitted by a mirror body depends
on its size as well as its surface temperature. 
Since the body equilibrates with the ordinary matter
surroundings, we can, to a good approximation,
define a common temperature for the body, $T' \simeq T$.
The amount of energy emitted by the body
is then \cite{dust}:
\begin{eqnarray}
E_{\text{emission}} &=& \int^{\infty}_0 4\pi^2 {R'}^2 Q_{R'} (R',\lambda) 
B_{\lambda} (T)
d\lambda
\nonumber \\
&=& \mu (T,R') \sigma T^4 4\pi R'^2
\end{eqnarray}
where
\begin{equation}
B_{\lambda}(T)=
\frac{2hc^{2}}{\lambda^{5}}\frac{1}{\exp\haak{hc/(\lambda k T)}-1} .
\end{equation}
For a perfect black-body emitter, $Q_{R'} (R',\lambda) = 1
\Rightarrow \mu = 1$. However for a finite sized solid body,
we have:
\begin{eqnarray}
Q_{R'} (R',\lambda) &\approx & 1, \text{ for } \lambda \ll R' \nonumber \\
Q_{R'} (R',\lambda) &\sim & R'/\lambda \text{ for } \lambda \gg R'
\end{eqnarray}
Thus, the net effect is:
\begin{eqnarray}
\mu (T,R') &\approx & 1 \text{ for } \ R' \gtrsim 
\text{few} \times 10^{-3}
\left( \frac{300 K}{T}\right) \text{ cm}
\nonumber \\
\mu (T, R') &\to & 0 \text{ for } \ R' \ll 10^{-3} \left(\frac{300 K}{
T}\right)\text{ cm}
\end{eqnarray}

Summarizing things, we have shown that a mirror matter
body embedded within ordinary matter 
(expected if $\epsilon < 0$) 
equilibrates and efficiently
emits mirror radiation provided that the body is greater than
about $10^{-3}$ cm in size. Of course, we cannot detect the
mirror radiation directly, however we can infer the existence
of an energy loss by looking for its thermal 
effect on the surrounding ordinary matter.
The point is that as the mirror body radiates its heat away
into mirror photons, the energy is replaced from the ordinary matter
surroundings which will become cooler as a result.

To examine the size of the effect, consider a spherical body of radius
$R'$ embedded within a much larger volume of ordinary matter.
In this case, the energy lost per unit time to mirror radiation 
is approximately, 
\begin{eqnarray}
Q_{\text{rad}} = \sigma T^4 4\pi R'^2 \ \text{ if } R' \gtrsim 10^{-3}\text{ cm}
\end{eqnarray}
As energy escapes, a temperature gradient in the surrounding 
ordinary matter is created, as heat is thermally conducted
replacing the energy lost into mirror radiation.
Thus, considering a shell of radius R surrounding the body ($R > R'$),
then the heat crossing this surface per unit time is:
\begin{eqnarray}
Q_c = \kappa \frac{\partial T}{\partial R} 4\pi R^2 .
\end{eqnarray}
In the steady state situation, $Q_c = Q_{\text{rad}}$,
which implies the temperature gradient,
\begin{eqnarray}
\frac{\partial T}{\partial R} = \frac{\sigma T^4 R'^2}{\kappa R^2} .
\end{eqnarray}
This means that the change in temperature of the ordinary matter
surrounding a spherical mirror body at a distance $R$, $\delta T(R)$,
is given by:
\begin{eqnarray}
\delta T(R) &=& \frac{-\sigma T^4 R'^2}{\kappa R}
\nonumber \\
& \simeq & -\left( \frac{T}{270 K}\right)^4 \left( \frac{R'}{5
\text{ cm}}\right)^2 \left( \frac{50\text{ cm}}{R}\right)\left( \frac{0.016
\ \text{J/(K.s.cm)}}{\kappa}\right)\ K
\nonumber
\\
.
\end{eqnarray}
This is clearly a significant thermal imprint which should be
detectable if large ($R' \gtrsim 5\text{ cm}$) fragments 
(or numerous small
fragments) exist close to the Earth's surface. In other words,
one could use a thermal imaging detector (or even an old thermometer!) 
as a mirror matter detector. But where should one look?

The chances of discovering a significant mirror matter deposit
in a random piece of earth would be expected to be quite low.
However, any mirror matter deposit near the Earth's surface
should be associated with the
impact of a mirror matter space-body.
Since the impact of a mirror matter body with the Earth has 
important observational differences compared with an ordinary body impact
it should be possible to identify candidate sites if they exist.
Remarkably, there is significant evidence for such
anomalous impact sites (see Ref.\cite{footyoon1,footyoon2,eros} for
a more detailed discussion):
\begin{itemize}

\item
On small scales ($D_{\text{SB}} \sim 5$ meters), there are relatively
frequent events, some of which have been quite well studied\cite{small}.  
At least one event of this type, the 2001 Jordan event, is particularly
interesting because the `body' was observed to hit
the ground. If caused by the impact of a mirror matter body, mirror
matter fragments should exist right on the ground and in the top surface
layers ($\sim$ cm depth). One would expect a sizable thermal imprint
at this impact site if caused by the impact of a mirror matter body.

\item
On large scales ($D_{\text{SB}} \sim 100$ meters), we have the famous
1908 Tunguska event\cite{tunguska}. In 
this case the body exploded in the atmosphere
at an altitude of about 8 km. If caused by a mirror matter
space-body, one would expect fragments on or near the surface, potentially
detectable by their thermal imprint.

\item
On even larger scales ($D_{\text{SB}} \sim 500$ meters), events should
be quite infrequent occurring on time scales much greater than
$10^3$ years. Yet, there is interesting evidence that such
anomalous impacts occur. One such event is 
the Edeowie glass
field in South Australia\cite{haines}. 
This event is characterized by a field of melted rocks (some of
which are actually melted in situ) on a flat desert plain over
an area of about $10^2 \text{ km}^2$. This event could be caused by a large
mirror body hitting the ground at cosmic velocity $v \sim 30$ km/s (which
is quite unlike the small Jordan event and Tunguska event).
In this case a large quantity of mirror matter should exist, but could
be buried some distance below the surface because of the
relatively large impact velocity (the stopping distance can be obtained
from Refs.\cite{footyoon1,eros} and is about 
$L \sim 10 (10^{-8}/\epsilon)^2 (v/30 \text{km/s})^4$ meters)\footnote{
For a large mirror matter deposit (which might be 
the case with the Edeowie glass field), another useful
way to infer its existence would be to use 
gravimeters to seach for local variations in the Earth's
gravity.
}.
\end{itemize}
The above anomalous impact sites should be a good place
to look for mirror matter via its thermal
imprint on the surrounding ordinary matter. Of course, once located
it can be further tested via the centrifuge technique\cite{centrifuge}.

In conclusion, we have shown that if the photon-mirror photon kinetic 
mixing parameter $\epsilon$ is in the range of interest, Eq.(\ref{zzz}),
mirror matter grains embedded within ordinary matter will thermally equilibrate 
with the ordinary matter environment. 
Consequently, a mirror matter grain will 
cool the surrounding ordinary matter by drawing in heat and 
radiating it away in the form of mirror photons. We have
shown that this cooling effect is significant enough to be 
detected for fragments larger than about a centimeter
in size (or numerous smaller fragments).
Thus, a mirror matter deposit could be located on Earth by its
thermal imprint on the ordinary matter surroundings. This
method, together with the centrifuge technique seems to
provide the most effective means to experimentally detect mirror
matter on Earth.

\vskip 1.3cm

\noindent
{\large \bf Appendix}

\appendix
\section{Calculation of Heat transfer}
If a mirror atom collides with an ordinary atom it can gain or 
lose kinetic energy, depending
on the initial velocities and the direction in which it scatters. 
It is convenient to express the gain
in kinetic energy of the mirror atom in terms of the initial 
velocities and scattering angles as follows.
Denote the initial velocity of the mirror atom 
(ordinary atom) as $v'$ ($v$), and the final
velocity of the mirror atom (ordinary atom) as $\tilde{v}'$ 
($\tilde{v}$). Define
x, y, z coordinates in velocity space as follows. 
Define the z-direction to be in the direction of $v'-v$.
The x-direction is chosen to be in the direction of the projection 
of $v$ on the plane orthogonal
to the z-direction. Finally, the y-direction is defined as the outer product
of the z and x-directions. Now define the spherical
coordinates $\theta$ and $\phi$ in the usual way. 
For any vector we define $\theta$ as the angle
with the z-direction and $\phi$ as the angle the vector projected 
on the x-y plane makes with the x-direction.
It follows from conservation of energy and momentum that if the final 
velocity of the mirror atom 
relative to the ordinary atom, $\tilde{v}'-\tilde{v}$, is in the
direction $\haak{\theta,\phi}$ then the gain in
kinetic energy is
\begin{equation}
\Delta E_{\text{kin}}\haak{v',v,\theta,\phi}=\frac{m}{4}\haak{1-\cos{\theta}}
\haak{v^{2}-v'^{2}}+ mv'v\sin\haak{\alpha}\sin\haak{\theta}\cos{\haak{\phi}}
\end{equation}
where $\alpha$ is the angle between the initial velocities.
To obtain the rate of heat transfer, we have to multiply this with 
the collision rate involving
mirror and ordinary atoms with the initial velocities $v'$ and $v$ 
and integrate over all initial
and final velocities. The velocity distribution of mirror atoms and 
ordinary atoms is
given by the Maxwell distribution. We 
define $n'\haak{v}$ and $n\haak{v}$ as giving the number
of mirror atoms respectively ordinary atoms per unit volume
and per unit volume in velocity space:
\begin{eqnarray}
n\haak{v}&=&n\haak{\frac{m}{2\pi k T}}^{\frac{3}{2}}\exp\haak{-\frac{m v^{2}}{2 k T}}\\
n'\haak{v}&=&n'\haak{\frac{m}{2\pi k T'}}^{\frac{3}{2}}\exp\haak{-\frac{m v^{2}}{2 k T'}}
\end{eqnarray}
Here $n$ ($n'$) is the number density of the (mirror) atoms. 
In terms of the above defined functions 
the amount of heat absorbed by the mirror fragment per unit volume 
and per unit time, $\frac{dQ}{dV}$, can be expressed
as:
\begin{equation}\label{qint}
\frac{dQ}{dV}=\int d^3 v'd^{3}v\sin\haak{\theta}d\theta d\phi \Delta E_{\text{kin}}\haak{v',v,\theta,\phi}
\frac{d\sigma}{d\Omega}\lhaak{v'-v} n'\haak{v'}n\haak{v}
\end{equation}
where $d\sigma/d\Omega$ is the elastic differential cross section. In
\cite{centrifuge} it is shown that for typical thermal velocities at
room temperature $d\sigma/d\Omega$ becomes approximately isotropic. We
can therefore put $d\sigma/d\Omega=\sigma/(4\pi)$ in Eq.\eqref{qint}.
The integrals in Eq.\eqref{qint} can be evaluated by substituting $v\rightarrow v'+u$,
and performing the $v'$ integration first. The result is:
\begin{equation}\label{rslt}
\frac{dQ}{dV}=
4\pi^{-1/2}m^{-1/2}k^{3/2}\sigma n'n\haak{\frac{T'+T}{2}}^{1/2}\haak{T-T'} .
\end{equation}

\vskip 1.3cm
\noindent
{\large \bf Acknowledgements}
\vskip 0.5cm
One of us (R.F.) would like to thank P. W. Haines for discussions
regarding the Edeowie glass field\cite{haines} and T. L. Yoon
for some comments on the paper.
 
\end{document}